\documentclass[useAMS,usenatbib]{mn2e}

\usepackage{graphicx}
\usepackage{amssymb}

\newcommand{\vecx}{\mathbf{x}}
\newcommand{\veck}{\mathbf{k}}

\newcommand{\fig}[1]{Figure \ref{fig:#1}}
\newcommand{\figs}[2]{Figures \ref{fig:#1} and \ref{fig:#2}}
\newcommand{\eqn}[1]{Equation \ref{eqn:#1}}
\newcommand{\eqns}[2]{Equations \ref{eqn:#1} and \ref{eqn:#2}}
\newcommand{\sect}[1]{Section \ref{sec:#1}}
\newcommand{\sects}[2]{Sections \ref{sec:#1} and \ref{sec:#2}}

\newcommand{\partiald}[2]{\frac{\partial #1}{\partial #2}}

\title[Non-Gaussian errors of BAO]{Non-Gaussian errors of baryonic acoustic 
oscillations}
\author[W. Ngan et al.]
{W. Ngan$^1$\thanks{Email: ngan@astro.utoronto.ca},
J. Harnois-D\'{e}raps$^{2,3}$\thanks{Email: jharno@cita.utoronto.ca},
U.-L. Pen$^2$\thanks{Email: pen@cita.utoronto.ca},
P. McDonald$^{4,5}$\thanks{Email: pvmcdonald@lbl.gov}
I. MacDonald$^{1,2}$\thanks{Email: macdonald@astro.utoronto.ca}\\
$^{1}$Department of Astronomy \& Astrophysics, University of Toronto, 50 St George Street, Toronto ON, Canada M5S 3H4\\
$^{2}$Canadian Institute for Theoretical Astrophysics, 60 St George Street, Toronto ON, Canada M5S 3H8\\
$^{3}$Department of Physics, University of Toronto, 60 St George Street, 
Toronto ON, Canada M5S 1A7\\
$^{4}$Lawrence Berkeley National Laboratory, 1 Cyclotron Road, Berkeley, CA, 
USA 94720\\
$^{5}$Brookhaven National Laboratory, Blgd 510, Upton NY 11375, USA
}
\begin{document}

\pagerange{\pageref{firstpage}--\pageref{lastpage}} \pubyear{2011}

\maketitle

\label{firstpage}

\begin{abstract}

We revisit the uncertainty in baryon acoustic oscillation (BAO) forecasts and 
data analyses. 
In particular, we study how much the uncertainties on both the measured mean dilation
scale and the associated error bar are affected by the non-Gaussianity of the non-linear
density field.
We examine two possible impacts of non-Gaussian analysis: (1) we derive the
distance estimators from Gaussian theory, but use 1000 N-Body simulations to
measure the actual errors, and compare this to the Gaussian prediction, and
(2) we compute new optimal estimators, which requires the inverse of the 
non-Gaussian 
covariance matrix of the matter power spectrum.
Obtaining  an accurate and precise inversion is challenging,
and we opted for a noise reduction technique applied on the covariance
matrices.
By measuring the bootstrap error on the inverted matrix, 
this work quantifies for the first time the significance of
the non-Gaussian error corrections on the BAO dilation scale.
We find that the variance (error squared) on distance measurements can deviate 
by up to 12\%
between both estimators, an effect that requires a large number of simulations 
to be resolved. 
We next apply a reconstruction algorithm to recover some of the BAO
signal that had been smeared by non-linear evolution, and we rerun the
analysis. We find that after reconstruction, the rms error on the
distance measurement improves by a factor of $\sim 1.7$ at low
redshift (consistent with previous results), and the variance
($\sigma^2$) shows a change of up to 18\% between optimal and
sub-optimal cases (note, however, that these discrepancies may depend
in detail on the procedure used to isolate the BAO signal).   
We finally discuss the impact of this work on current data analyses.
 
\end{abstract}



\section{Introduction}

Ever since acoustic peaks were detected in the cosmic microwave
background \citep{miller} and galaxy surveys \citep{sdss}, much effort
has been devoted to use baryon acoustic oscillations (BAO) as standard
rulers to estimate cosmological distances with precision \citep{bao,
  bg2003, se03, baoreview, pat07}.  BAO signals are manifested as a
wiggly feature in the matter power spectrum, and precise measurements could shed
light on the dynamics of dark energy.  
The accuracy and precision of such measurement depends directly on the covariance
matrix of the matter  power spectrum, which is straightforward to compute for
a Gaussian density field using Wick's theorem.  However, it is known
that non-Gaussianities are significant in that covariance matrix \citep{Meiksin, scoccimarro,rh2005, rh2006, jap} 
and may have an impact on cosmological distance measurements. 
A non-linear model is needed for a non-Gaussian calculation, 
hence some criterion for the accuracy of such effects is needed
to quantify the precision of the measurement.
In particular, an optimal BAO measurement should in general incorporate a 
proper error 
weighting of the data,
which involves the inversion of the full covariance matrix. 
The accuracy criterion must therefore be based on the confidence we have on the inverted matrix.

The estimator constructed in most forecasts and data analyses follow the prescription of \citet{se07}, which 
is a procedure that constructs both the estimator of the BAO scale and the estimator of its uncertainty under the Gaussian assumption.
The method was originally cross-checked with a $\chi^{2}$ analysis of $51$ 
simulations, including a jackknife sub-sampling  of the N-body data,
and showed good agreement.
The actual deviation between this proposed estimator and the optimal estimator 
constructed in this paper is indeed small, 
therefore consistent with the work of  \citet{se07}. 
In this paper, we aim at improving on the accuracy of that method, and present 
a first significant detection 
of the effect of non-Gaussian errors on the measurement of the BAO scale 
(the effect of non-linear evolution, in the form of erasure of the wiggle 
signal, has been well-studied of course \citep{se07}).

Some data analyses \citep{hutsi, sdss7, sdss, cole} also improved on the original method 
by constructing a non-Gaussian estimator for the BAO dilation scale uncertainty.
This is typically done by modeling the non-linearities in the density fields with mock catalogs,
which are produced from log-normal densities \citep{sdss7,  colesjones},
 $2^{nd}$ order perturbation theory \citep{hutsi}, halo models \citep{sdss}, 
etc. 
Such techniques all attempt to increase the robustness of the analysis 
by taking into account the coupling between the Fourier modes. 
As mentioned above, an optimal analysis must be based on a reliable
inverted covariance matrix, and the accuracy of the inverse matrix constructed 
from mock catalogs is yet to be demonstrated.
The covariance matrix in power spectrum is a four point function
that relates pairs of wave numbers. 
The error on this covariance consists of pairs of these pairs, and is
indeed difficult to quantify.  Without this metric, however, one does not
know the significance of a non-Gaussian computation.
In addition,  \cite{jap2} have found significant departures between
the covariance matrix  constructed from Lagrangian perturbation theory and that obtained from 
their $5000$ simulations.
Also unknown is the accuracy of log-normal densities at
modeling the true covariance matrix and its inverse.
Other analyses \citep{WiggleZ} treat the mode coupling as
coming exclusively from the survey selection function, following the widely used
FKP \citep{FKP} prescription. This specific coupling effect can be reduced 
with other choices of power spectrum estimators, like that presented in 
\citet{tegmark}.
In both cases, however, the non-linear mode coupling is not modeled.

When it comes to the impact on the BAO dilation scale, 
a Gaussian treatment of the data yields a sub-optimal estimation of
the mean, and the error bars obtained that way are systematically
biased, usually on the low side.  In the limit where the sample
is large enough, the value of the mean estimated in that fashion
does converge to the ``true'' mean, but the estimated error bars never
capture the correlation that occurs in the non-linear regime.  
Many analyses attempt to correct for this bias with Monte Carlo simulations, 
however this effect is very small and takes a high accuracy and precision
to observe.

In the non-Gaussian case, however, an inversion of the covariance matrix is 
required, 
hence it cannot be singular. Consequently, the convergence of our measured
matrix depends on the binning, and the inversion increases the noise even more.
For an adequate resolution on all the scales relevant for BAO analyses, 
the number of simulations required to obtain statistically significant 
conclusions
can be large. In the past, different groups used drastically different
numbers of simulations: \citet{se05} used 51, \cite{rh2005} used 400,
and \cite{jap} used 5000; we use  1000 N-body simulations  in this work.  
In order to invert a $N\times N$ covariance matrix, one needs {\it at least} $N$ simulations to make
the matrix non-singular.  It is also generally thought that to achieve
convergence on each element, we need of the order $N^{2}$ simulations.
Even then, the level of accuracy is not clear, and 
to make a significant claim about non-Gaussian effects, 
one needs to know the uncertainty on the inverted matrix, 
which we measure from a bootstrap re-sampling, and to propagate 
the error onto the BAO scale.

To address this convergence issue, we also apply a noise reduction technique 
before the inversion:
we factorize the covariance matrix with an Eigenvector decomposition, and keep 
only the principal component.  
This factorization is repeated at each of the bootstrap samplings,  
which allows us to draw robust conclusions on the convergence of our results.

Given the fact that the precision of the inverse covariance matrices used in 
analyses has never been demonstrated, the measurement of the mean and of the 
error on the mean found on the literature are most likely not optimal.
Measuring an optimal BAO scale in actual data is complicated in many aspects, 
such as the fact that the Universe is not periodic,
and that surveys have selection functions. It would nevertheless involve a 
covariance weighted measurement,
which is not included in the prescription given by \citep{se07}.
If one could improve the measurement of the power spectrum covariance, however
i.e. from N-Body simulations, it would be possible to measure a more robust and  more accurate
{\it uncertainty} on the sub-optimal mean, compared to the original claim.  
The difference in performance between these two BAO dilation scale estimator 
is still an unmeasured  quantity. In this paper, we first attempt to address 
this question by
comparing three different analysis scenarios:

\begin{enumerate}
\item The first case we consider is an attempt at measuring a correct
  error bar on a sub-optimal mean of the BAO dilation scale. 
  Even if we know that the Universe is non-Gaussian, it is still possible to
  treat it as Gaussian, i.e. not use an optimally weighted sum when
  estimating the mean, even though the power spectrum itself is non-linear. 
    Doing so, we must keep in mind that the
  measurements are non-optimal, and that the naive Gaussian error bars
  are most likely too small compared to the ``true" error.  However,
  given the fact that we can measure a full covariance matrix from
  N-body simulations, we can get a better estimate of the error bars
  on that sub-optimal mean by treating the original covariance matrix
  as noisy and by performing an appropriate inverse covariance weighting.  
  From now  on, we refer to this case as the ``sub-optimal'' estimator of the
  BAO error.  This approach is commonly used to obtain ``Monte-Carlo''
  error bars, and exactly what is measured by the forecasting prescriptions 
  mentioned above.
  We call this approach ``sub-optimal'' because the
  errors bars could still be further reduced by improving the
  estimator of the mean BAO scale using the non-Gaussian model.  
  It provides a quick estimate of the magnitude of non-Gaussianity, and a 
  simple way of
  scaling errors bars obtained from Gaussian analysis, which is often
  used in  surveys.  

\item The second case, dubbed ``optimal estimator'', is the best quadratic
  analysis one can possibly do -- knowing that the Universe is
  non-Gaussian, we treat it as is, measuring a fully non-linear power
  spectrum covariance matrix and performing an optimally weighted sum
  to estimate both the means and the uncertainties on the parameters.
  Given the fact that we rely on a large number of N-Body simulations,
  that our volume is periodic, and that we have a high signal to
  noise, then both our estimators are truly optimal in the least squares sense.
  \footnote{In the case of a data analysis,  however, the optimal measurement 
  of the covariance matrix is complicated by the fact that the underlying 
  matrix $C(\mathbf{k},\mathbf{k'})$ is six-dimensional, and non-isotropic in the 
  sense that 
  pairs of mode separated by smaller angles are more correlated.  
  On top of that, the observed quantity is convoluted in six-dimensional
  with pairs of survey selection function \citep{Harnois}. 
  For these reasons, we find solutions that apply to periodic volumes, 
  and we leave it for future work the extraction of optimal estimator
  in more complex data.}

\item Many times in the literature \citep{WiggleZ, tegmark,P01}, the above case is
  modified by replacing the non-linear covariance matrix by a Gaussian
  one. This effectively treats the power spectrum measurements as
  Gaussian, even though the data are correlated.  Under the widely used FKP
  \citep{FKP} approximation, for example, the only source of mode
  coupling comes from the convolution with the survey selection
  function; it thus considers the underlying  covariance matrix to be
  diagonal.  For reasons mentioned above, the error bars obtained this
  way could be systematically underestimated.
\end{enumerate}

We did not address the question of shot noise, however, which
is also non-Gaussian in nature.
In the case of surveys that address some of the non-Gaussianities with
mock catalogs, the error correct bars should lie somewhere between case (i)
and case (ii) (while some approach case (iii)), depending on how close
to optimal the measured power spectrum is, and how well the catalogs
model the non-linear dynamics.  \citet{jap} showed that the difference in
estimator between cases (ii) and (iii) is very small (i.e., with
optimal weighting, errors near the pure-Gaussian errors are achieved),
but this is not the complete story, especially when dealing with
current Gaussian or sub-optimal data analyses.  The question we
address is the following -- by how much does the error bars on the
{\it least robust} BAO dilation scale (case (iii)) differ from a
correct calculation based on an optimal covariance matrix  and
properly inverse covariance weighted (case (i))?

Finally,  we go one step further and repeat the measurements of non-Gaussian 
effects on reconstructed density  fields.
We apply a density reconstruction algorithm \citep{recon}
that was developed to improve the BAO signal at late times, which is
partially lost due to non-linear coupling between Fourier modes
\citep{recon2, recon3, seo09}.  Other approaches have been explored to
recover some of the Fisher information lost in gravitational collapse
\citep{Goldberg, zhang, zhang2, Neyrinck, seo10}, but it was not
verified how these methods propagate to constraints on cosmological
parameters.  The full BAO analysis is indeed sensitive to this
intermediate stage, but in a non-Gaussian treatment, the interplay
between the off-diagonal elements of the covariance matrix and the
derivatives is quite subtle.  We thus set forth to test quantitatively
how this reconstruction algorithm affects the BAO dilation error, for
the three analysis cases mentioned above.

The paper is organized as follows. 
In \sect{background} we briefly review how BAO dilation measurements can 
constrain dark energy.
In \sect{nbodysimulations} we describe our set of N-body simulations and 
the reconstruction algorithm.
In \sect{powerspectrumanalysis} we discuss how to extract both the Gaussian and
the  non-Gaussian covariance matrices. 
In \sect{Eigen} we describe the Eigenvector decomposition, while
in \sect{parameterestimation} we present 
the Fisher matrix formalism and the estimators of the BAO  dilation  
uncertainty for the three analysis cases. 
Finally, in \sect{resultsanddiscussion} we present and discuss our results.


\section{Background}
\label{sec:background}

\subsection{Baryon acoustic oscillations}

The matter clustering we observe today is the result of tiny inhomogeneities
set during inflation in the early Universe \citep{inflation}. Over time, matter
collapsed 
gravitationally into over-dense regions that eventually evolved into large scale
structures. As long as the perturbations are small, the structure growth 
equations can be linearized. We can therefore understand the evolution of 
inhomogeneities by considering perturbations one at a time.

In the early Universe, matter and photons are coupled together as a single 
fluid via Thomson scattering. 
Due to radiation pressure, photons begin to disperse away from over-dense 
regions, pushing the baryons alongside.  
Dark matter, on the other hand, only interacts weakly with that fluid, via 
gravity, and does not respond efficiently to the photons' push. The 
perturbation eventually grows into a state where the initial clump of dark 
matter gets surrounded by a spherical ripple of 
baryon-photon fluid, which expands at the speed of sound $c_s\simeq c/\sqrt{3}$.

At about $z\sim1000$, when photons decouple from matter, they no longer push 
the baryons. The speed of sound in the fluid drops abruptly, and the BAO ripple 
stops moving and freezes out. Eventually, dark matter also responds 
gravitationally to this over-dense region of baryons. The result of this 
initial 
point-perturbation is a smooth clump of matter with a spherical shell of density
enhancement at the {\it sound horizon},  about 150 Mpc away  from the center. 
The complete final field is, by Huygens's principle, the superposition of 
similar 
spherical ripples of density enhancement from the initial perturbations at all
points. Therefore, we do not observe these spherical ripples directly, 
but measure them statistically in the mass auto-correlation function.

Using galaxies from the  Sloan Digital Sky Survey  as tracers of the matter 
distribution, 
an excess correlation at 150 Mpc apart has been observed \citep{sdss}.
Although the BAO in the correlation function is intuitive, 
we are interested in the BAO power spectrum
$P(k)$ which is related to $\xi(r)$ by a transformation
\begin{equation}
 \xi(r) = \frac{1}{2\pi^2}\int k^2 P(k) \frac{\sin(kr)}{kr} dk.
\end{equation}
The correlation peak in real space is manifested as a series of wiggles in 
Fourier space.
This oscillatory feature is even more robust against observational 
contaminations \citep{se03}.

\subsection{Dark energy constraints}
\label{sec:darkenergyconstraints}

The accelerating expansion of the Universe is widely blamed on dark energy, a 
mysterious entity which contributes to more than 70\% of the energy content in 
the Universe. Dark energy can be described by an  equation of state
\begin{equation}
 P = w c^2 \rho
\end{equation}
relating its pressure $P$ and density $\rho$. 
A common parameterization of $w$ is 
\begin{equation}
 w(z) = w_0 + \frac{w_1z}{1+z}
\end{equation}
where $w_0$ is the value in the present day. The Friedmann equation now reads
\begin{equation}
\begin{array}{rcr}
 H^2(z) = H_0^2 \left[ \Omega_m(1+z)^3 + \Omega_r(1+z)^4 + \Omega_k(1+z)^2 \right. \\
\left.+\Omega_\lambda \exp\displaystyle\left(3\int_0^z \frac{1+w(z)}{1+z}dz\right) \right]
\end{array}
\label{eqn:friedmann}
\end{equation}
where the terms on the right hand side are the energy contributions of matter, 
radiation, curvature, and dark energy respectively.
For an object of a fixed co-moving size  $s$, its projections across
and along the line of sight are given by
\begin{equation}
 s_{||} = \frac{c\Delta{z}}{H(z)}
\end{equation}
\begin{equation}
 s_\perp = (1+z) D_A(z) \Delta{\theta}
\end{equation}
respectively.  $\Delta z$ and $\Delta\theta$ are the redshift span and angular size of 
the object on the sky, and
\begin{equation}
 D_A(z) = \frac{c}{1+z} \int_0^z \frac{dz^\prime}{H(z^\prime)}
\end{equation}
is the angular diameter distance to its center. If this object has a fixed co-moving size,
it then acts as a standard ruler, 
where measurements of $\Delta z$ and $\Delta\theta$ give estimates for $H$ and 
$D_A$, respectively. This, in turn, provides constraints for $w(z)$ via \eqn{friedmann},
given an adequate redshift sampling.

In this paper, we restrict ourselves to an idealized isotropic universe
with no observational distortions, such that the standard ruler is given by 
the sound horizon, which has size $s=s_{||}=s_\perp$.  
Therefore, the fractional errors on these quantities can be used to constrain 
$H(z)$ and $D_{A}(z)$. To estimate the error on $s$, we construct a Fisher 
matrix that 
propagates the correlated uncertainty measured in our simulated power spectra.


\section{Simulations and Density Reconstruction}
\label{sec:nbodysimulations}


We run a total of 1000 simulations using a particle-particle-particle-mesh 
(P$^3$M) N-body code {\sc 
cubep3m}\footnote{http://www.cita.utoronto.ca/mediawiki/index.php/CubePM},
the successor to {\sc pmfast} \citep{pmfast}. 
Each simulation has $N=256^3$ dark matter particles in a periodic cube of 
$L=600\ h^{-1}$Mpc on one side. The initial condition of each simulation at 
$z=100$ is 
produced by a Gaussian random field, which is characterized by an initial 
transfer function generated by {\sc CAMB}\footnote{http://camb.info/}. 
For this work we use the following cosmological parameters: $\Omega_m=0.279$, $
\Omega_b = 0.044$, $\Omega_\Lambda = 0.721$, $h = 0.701$, $n_s = 0.96$, $
\sigma_8 = 0.817$.
We output the particle positions and velocities  at redshifts 0.5, 1.0 and 2.0, 
and then the particles are assigned to a density field $\delta(\vecx)$ on a 
$512^{3}$ grid using the cloud-in-cell algorithm \citep{cic}.

To understand the density reconstruction algorithm, it is helpful to look at 
the process by which
initial conditions are generated in a typical N-body simulation.
A Gaussian random field $\delta(\veck)$ in 
Fourier space can be constructed by generating a field of  Gaussian random 
numbers 
whose variance is determined by an input power spectrum. 
Under the Zel'Dovich approximation \citep{zeldovich}, we then compute a 
displacement field
\begin{equation}
    \mathbf{s}(\veck) = -\frac{i\veck}{|\veck|^2} \delta(\veck)
    \label{eqn:displacementfield}
\end{equation}
which, when Fourier transformed back into real space, can be applied to 
displace 
a uniformly distributed set of particles. In addition, the density field allows
us to calculate the 
gravitational potential, whose gradient gives the particles' initial 
velocities. 
We solve for these initial conditions at $z=100$, where our 
simulations begin.


The reconstruction algorithm developed by \citet{recon} essentially uses the 
$\delta(\vecx)$ output to calculate a displacement field, and subtracts the 
displacements from the particle positions. This is indeed very similar 
to the procedure that generates initial conditions  described previously,
except that the displacements are subtracted from the particles' positions, 
instead of added. 
The algorithm is the following, as was neatly summarized by \citet{recon2}. 

\begin{enumerate}
	\renewcommand{\theenumi}{(\arabic{enumi})}
	\item Calculate the density field $\delta(\vecx)$ using particle 
                       positions, 
              and then transform it into Fourier space $\delta(\veck)$. 
 
	\item Calculate the displacement field $\mathbf{s}(\veck)$ in Fourier 
              space using the Zel'Dovich approximation, where
          \begin{equation}
            \mathbf{s}(\veck) = -\frac{ik}{|\veck|^2}\delta(\veck)F(\veck),
            \label{eqn:zeldovich}
	  \end{equation}
           and $F(\veck)=\exp[-(kR)^2/2]$ is a smoothing function of scale $R$.

	\item Transform the displacement field back into real space. Subtract 
              this displacement from the positions of the simulation particles 
              and calculate the new density field $\delta_d(\vecx)$.
	
	\item Repeat the previous step, but applying the  displacement field 
               onto a set of uniformly distributed particles instead of 
                simulation output.  Calculate the density field 
                 $\delta_u(\vecx)$ from these displaced particles.
	
	\item The ``reconstructed'' density field $\delta^\prime(\vecx)$ 
                  is given by the difference between the above density fields:
	\begin{equation}
		\delta^\prime(\vecx) = \delta_d(\vecx) - \delta_u(\vecx).
	\end{equation}

\end{enumerate}

\begin{figure}
    \centering
    \includegraphics[width=3.5in]{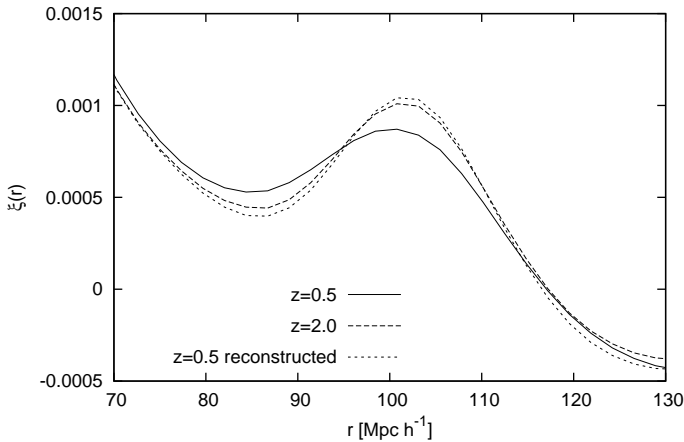}
    \caption{Correlation function $\xi(r)$ at redshifts $z=0.5$ and $z=2.0$, as 
well as reconstruction (smoothing scale $R=10\ h^{-1}$Mpc) at $z=0.5$. The 
functions are rescaled by the square of the growth factors so that the acoustic
peak locations match the $z=0.5$ case.}
    \label{fig:xi}
\end{figure}

The correlation functions $\xi(r)$ before and after reconstruction 
(using smoothing scale $R=10\ h^{-1}$Mpc) are shown in \fig{xi}. 
In \sects{effectsofreconstruction}{resultsanddiscussion}, we quantify the 
effect that reconstruction has on distance error estimates. 


\section{Power spectrum analysis}
\label{sec:powerspectrumanalysis}

\subsection{Matter power spectrum}

In the presence of anisotropy (eg. redshift 
distortions), the power spectrum $P(k,\mu)$ takes in an angular dependence
$\mu \equiv \cos\theta$. 
In an isotropic universe, however, it is only a function of the scale and is 
defined as
\begin{equation}
    P(k) = \langle |\delta(\veck)|^2 \rangle
\end{equation}
where the angled brackets denote the average of all $\veck$ modes such that 
$|\veck|
=k$. To obtain $P(k)$ from our N-body simulations, we assign particles onto a 
density grid $\delta(\vecx)$, Fourier transform it into $\delta(\veck)$, and 
take the averages over the spherical shells of radius $k$ in Fourier space, 
using the nearest-grid-point scheme. 

\begin{figure}
    \centering
    \includegraphics[width=3.5in]{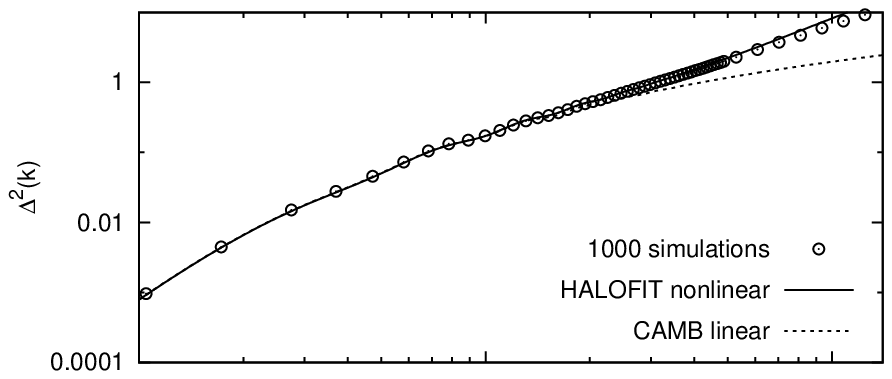}
    \includegraphics[width=3.5in]{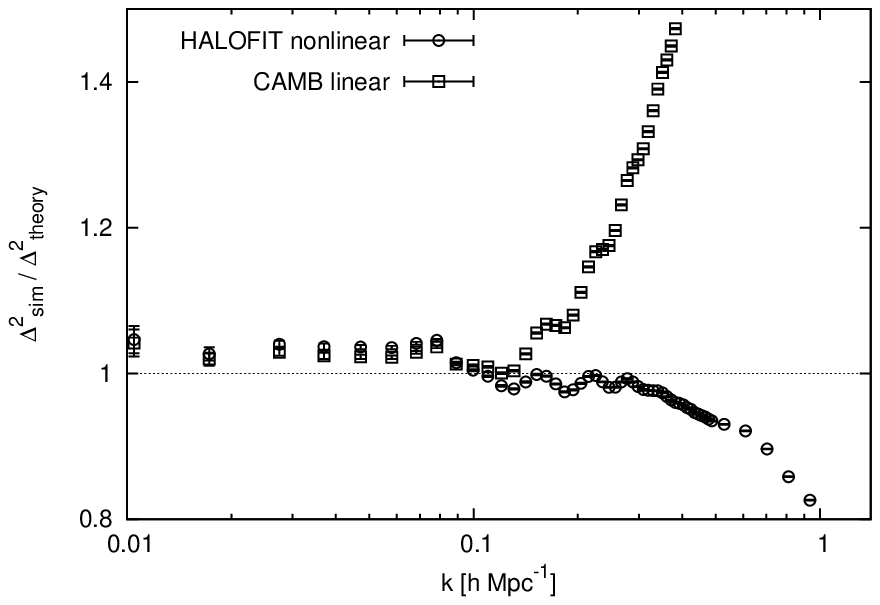}
    \caption{Dark matter power spectrum at $z=0.5$ produced by averaging 1000 
simulations. {\it Top panel}: The simulated dimensionless power spectrum 
$\Delta^2(k)=P(k)k^3/2\pi^2$
compared against calculations by CAMB, where 
``linear'' is from the linear $\Lambda$CDM theory, and ``nonlinear'' is based 
on HALOFIT 
\citep{halofit} numerical calculations. {\it Bottom panel}: The simulation 
power spectrum divided by linear theory
and HALOFIT power spectra. The first 47 modes of our spectrum are binned 
linearly, while the 
others are binned logarithmically. The error 
bars in each plot are the errors on the mean.
We note the few percent discrepancy between the output and linear theory on 
large scales,  which is caused by overly large time steps taken
by the simulator at high redshifts,
and will be corrected in further simulations.
However we have ran convergence test on the starting redshift of the
covariance matrix and found that a starting redshift of $z = 100$ produces the most accurate
matrix both at high and low redshifts. 
} 

    \label{fig:powerspectrum}
\end{figure}

Because our goal is to measure both a covariance matrix and a Fisher matrix 
from a finite number of realizations, 
the binning must be chosen carefully.
On one hand, we need to be maximally sensitive to the BAO signal, 
hence it is important to resolve many wiggles. Otherwise, our analysis would
under-sample the rapidly oscillating signal and our results would be less 
robust.
On the other hand, we need to limit the number of matrix elements 
to address the issue of convergence. 
We thus opt for mixed binning, which is linear for $k<0.5\ \rmn{Mpc}^{-1}h$
($\Delta k\sim 0.01\ \rmn{Mpc}^{-1}h$)\footnote{The linear bin width is not exactly $2\pi/L=0.0105\ \rmn{Mpc}^{-1}h$
where $L$ is the box size because the bins have been corrected for averaging with the nearest-grid-point scheme.
This correction is significant for small $k$ where the number of modes that contributes to the average
is small.}
and logarithmic for larger $k$ modes ($\Delta \log k = 0.062$)
. We end up with 54 bins, which translates into 2916 matrix elements, 
and we resolve the first seven BAO wiggles. 
\fig{powerspectrum} shows the average $P(k)$ over 1000 simulations. 

\subsection{Covariance matrix}
\label{subsec:covmatrix}

Given our series of $P(k)$ measurements, we 
calculate the covariance matrix between each data point as 
\begin{equation}
    C(k,k^\prime) = \frac{1}{n-1}\sum^n_{i=1}\left[P_i(k)-\langle P(k)\rangle
\right]\left[P_i(k^\prime)-\langle P(k^\prime)\rangle\right]
    \label{eqn:covariance}
\end{equation}
where $P_i(k)$ is the power spectrum of the $i$th simulation, and $n=1000$ is 
the number of simulations we have.

The power spectrum covariance matrix is best visualized as a cross-correlation 
coefficient matrix 
\begin{equation}
	\rho(k,k^\prime) = \frac{C(k,k^\prime)}{\sqrt{C(k,k)
                     C(k^\prime,k^\prime)}}.
	\label{eqn:corrmatrix}
\end{equation}
This definition normalizes the covariance matrix by the diagonal components so 
that they are all unity. This is consistent with the fact that a given $k$ mode 
is always perfectly correlated with itself. Therefore, $\rho(k,k^\prime)$ 
ranges 
from $-1$ (perfect anti-correlation) to +1 (perfect correlation), where 0 is no 
correlation  at all.

\begin{figure}
    \centering
    \includegraphics[width=3.5in]{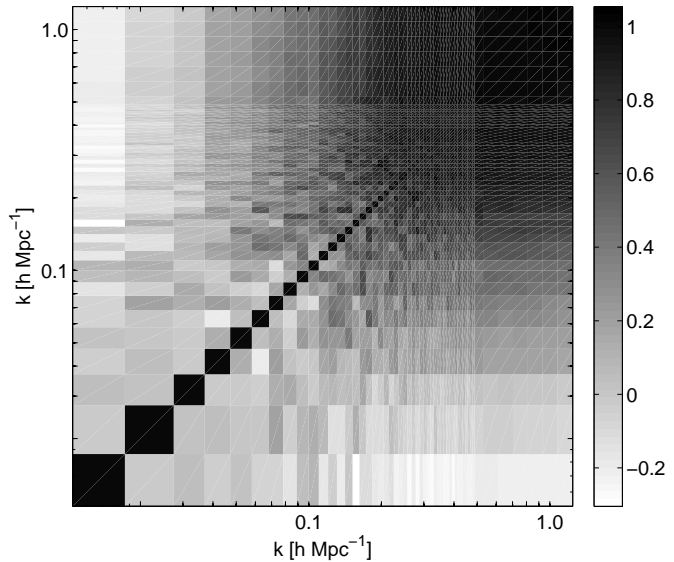}
    \caption{Correlation matrix (\eqn{corrmatrix}) at $z=0.5$ produced by 
averaging 1000 simulation power spectra as shown in \fig{powerspectrum}. 
The dark points on the diagonal and at the large-$k$ regime indicate positive 
correlation, as we expected. 
As described in \sect{Eigen}, this figure shows a binning independent 
version of the correlation matrix.}
    \label{fig:corrmatrix}
\end{figure}

\fig{corrmatrix} shows the correlation matrix corresponding to the 1000 power 
spectra shown in \fig{powerspectrum}. 
We observe that the largest $k$-modes in the simulation box exhibits an 
anti-correlation
of about $20\%$ with the small scales, which themselves are highly correlated. 
The systematic anti-correlation has very little impact on the final measured 
quantities.
We find that neither cutting off these three fundamental modes nor suppressing
those negative covariance values to zero at late redshifts affect our results. 
This is not surprising because, as we show in the
following sections, the lowest-$k$ modes contain very little Fisher information, while the
largest-$k$ modes usually do not contribute much to our distance errors. 
Therefore, the presence of this anti-correlation has no influence on our 
conclusions.

\begin{figure}
	\centering
	\includegraphics[width=3.5in]{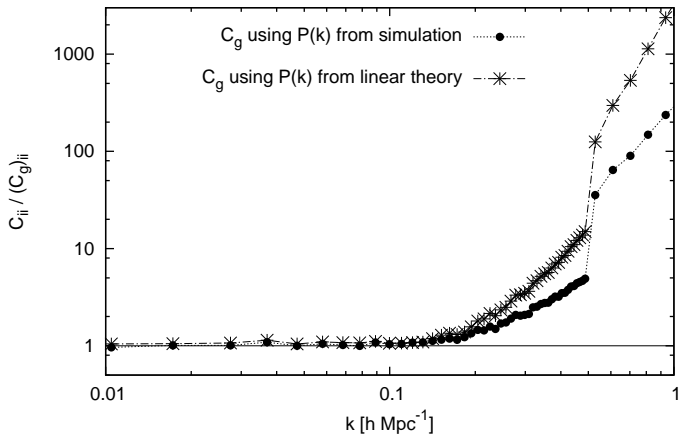}
	\caption{The ratio between the diagonal components of the non-Gaussian 
        ($C_{ii}$)
	and Gaussian ($(C_g)_{ii}$) covariance matrices, at $z=0.5$. 
        For $C_g$ we evaluated in 2 ways. Using \eqn{covariancegauss}, the 
        $P(k)$ can 
	either be averaged over our set of simulations, or simply computed 
        from linear theory
        using CAMB. For either case, this plot shows that even the
	diagonal components of the non-Gaussian covariance matrix are not 
        Gaussian,
	as the ratio clearly departs from unity with higher $k$. 
        The sharp increase at $k \sim 0.5\ \rmn{Mpc}^{-1}h$ is caused by the 
        logarithmic binning,
        in which each bin suddenly accumulates the departure contributions from
        many more Fourier modes.
        As we will justify in the results discussion later, we do not consider 
        modes where $k>0.5$ Mpc$^{-1}h$.
}
	\label{fig:gaussratio}
\end{figure}

In the three analysis cases studied in this paper,
we work with two kinds of covariance matrices. The {\it non-Gaussian} 
covariance 
matrices are computed using \eqn{covariance}, while the
{\it Gaussian} covariance matrices \citep{gausscov} are given by
\begin{equation}
    C_g(k,k^\prime) = \frac{2 P(k)P(k^\prime)}{N_k} 
\delta_{kk^\prime}
    \label{eqn:covariancegauss}
\end{equation}
where $N_k$ is the number of modes that contributed to the $k$ bin (counting
the real and imaginary parts separately), and $P(k)$ can be calculated from 
linear theory.
Indeed the 
Kronecker-delta symbol $\delta_{kk^\prime}$ ensures that the Gaussian 
covariance 
matrix is diagonal, and assumes that all the $k$ modes are uncorrelated. 
In a real survey, when only one Universe can be observed, it is much harder to 
calculate the 
correlation between different $k$ modes.
As shown in \fig{corrmatrix}, the actual 
covariance matrix is clearly not diagonal as we approach the non-linear regime. 
Furthermore, as shown in \fig{gaussratio}, even
the diagonal components of the non-Gaussian covariance matrix are not
Gaussian. The Fourier modes on small scales are coupled both with neighboring 
scales, and across different directions on a given scale.
In this context, the best estimate of the true covariance matrix is 
obtained  from an ensemble of N-body simulations, and the resulting non-Gaussian
treatment of the data is our best shot at estimating both the mean 
and the error on the mean. The replacement of the non-Gaussian covariance 
matrix by a 
Gaussian one is often used as a  shortcut, but one should keep in mind 
that the results in general may not be reliable.

\subsection{Fisher information function}
As previously considered by \citet{rh2005, rh2006}, a useful quantity that can
be 
derived from the covariance matrix is the cumulative Fisher information function
\begin{equation}
    I(k_j) = \sum^{k_j}_{k,k^\prime} C_{norm}^{-1}(k,k^\prime)
    \label{eqn:info}
\end{equation}
where $C_{norm}$ is the $j\times j$ sub-matrix of $C(k,k^\prime)$ (or $C_g$) 
with 
$k,k^\prime\leq k_j$, further normalized by $\langle P(k)\rangle
\langle P(k^\prime)\rangle$. In other words,
\begin{equation}
    C_{norm}(k, k^\prime) = \frac{C(k,k^\prime)}{\langle P(k)\rangle \langle 
P(k^\prime)\rangle} \qquad (k,k^\prime \leq k_j\ \rmn{only}).
\label{eqn:C_norm}
\end{equation}
Here $I(k)$ describes the information contained in the amplitude of
the power spectrum up to a scale $k$. Although the form of $I(k)$ seems like a 
mathematical contrivance,
it indeed has physical significance in weak lensing \citep{lupen}.

\begin{figure}
    \centering
    \includegraphics[width=3.5in]{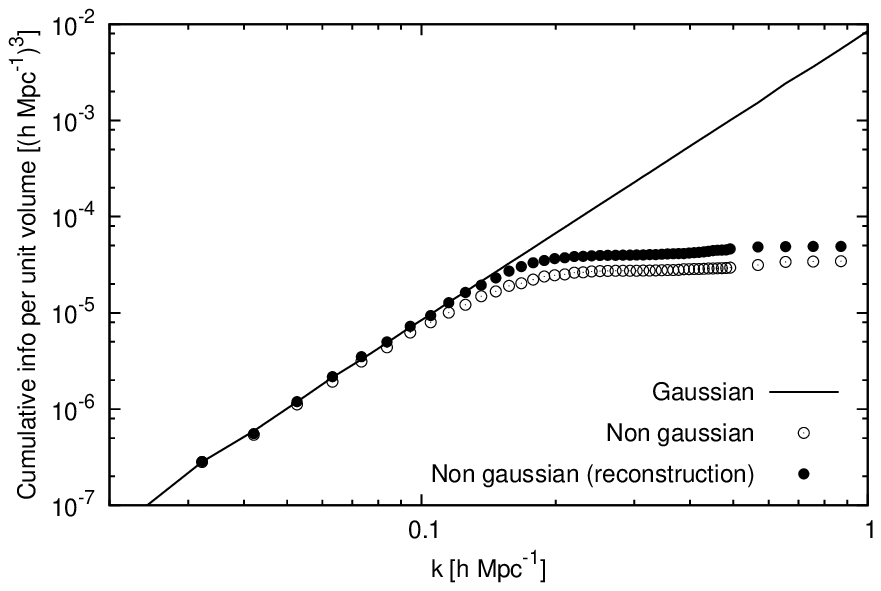}\\
    \includegraphics[height=1.48in]{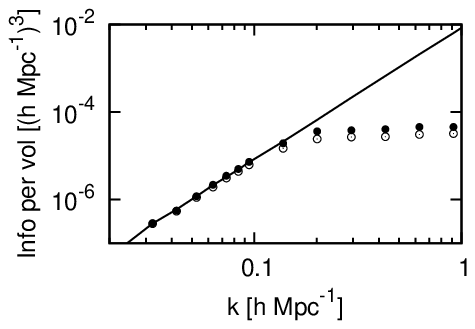}
           \includegraphics[height=1.48in]{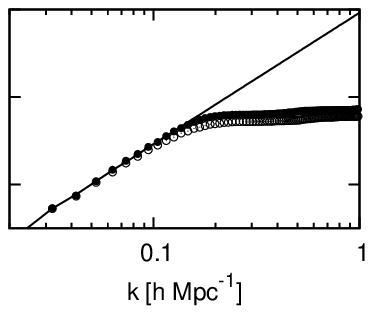}
    \caption{{\it Top panel}: Cumulative information (\eqn{info}) per unit 
                volume at $z=0.5$. The filled circles were obtained
                by using both the reconstructed power spectrum and the reconstructed
                 covariance matrix in \eqn{C_norm}. {\it Bottom panel}: Same plots as the top, 
                   but with different binning schemes to show that 
                 the plateau of information is not a binning artifact. Similar 
               plots can also 
               be found in \citet{rh2005, rh2006}.
   }
    \label{fig:information}
\end{figure}

\fig{information} shows the Fisher information function using our set of 
simulations. It also compares the information that can be obtained by a 
Gaussian 
and a non-Gaussian covariance matrix. At small $k$, gravity is still linear, 
so the non-Gaussian information is similar to the Gaussian case, as also seen 
in \fig{corrmatrix}. 
At $k\gtrsim 0.3\ \rmn{Mpc}^{-1}h$, however, the information flattens out into 
a so-called
``trans-linear plateau'', which is consistent with previous studies by 
\cite{rh2005} and \cite{jap}.
This corresponds to the scale where the $k$ modes become correlated. 

Qualitatively, the Fisher information function tells us how much information 
can 
be extracted from a given $k$ mode. At small $k$, each mode is independent, so 
more information can be extracted by measuring more modes. At large $k$, the 
Fisher information function flattens, meaning very little information can be 
extracted by measuring extra modes. Indeed, as seen from \fig{information},
one can extract orders of magnitude more information when using a 
Gaussian covariance matrix instead of the true covariance matrix. 
This information, however, does not exist in the data. Consequently, an 
analysis using a Gaussian estimator would most likely underestimate the error 
bars for any measurements beyond the trans-linear scale 
($k\gtrsim 0.3\ \rmn{Mpc}^{-1}h$).

Before proceeding to the construction of the Fisher matrix and of the three 
estimators, 
we reduce the noise that exists in our covariance matrices, as described in the
following section.

\section{Noise reduction}
\label{sec:Eigen}
As mentioned in the introduction, to achieve convergence on the  $N^{2}$ 
elements of the covariance matrix is not an easy task. 
Among the possible approaches, the brute force way 
(i.e.  running a very large number of N-body simulations)
is probably the least affordable in terms of computational resources. 
In this section, we apply a noise reduction technique that reduces to $2N$ the 
number of elements one needs to extract. 

Our technique is based on the assumption that the cross-correlation coefficient matrix  (see \fig{corrmatrix})
can be parameterized as the sum of a diagonal and off-diagonal components,
and we put a strong prior on the latter: it must be expressed as a sum over the outer products
of the principal Eigenvectors\footnote{Because the off-diagonal elements are monotonically increasing
both towards higher $k$- and $k'$-modes, the number of significant
Eigenvectors is expected to be small. In contrast, a matrix which would be
strong when close to the diagonal, but decreasing when moving away,
would require a larger number of vectors.}.
The diagonal component is then simply obtained such as
each diagonal element in the final matrix is equal to 1.

To extract the Eigenvectors, we perform an iterative eigenvalue decomposition of the
off-diagonal component and keep only the dominant terms.
We start by modeling the full matrix as the sum of an identity matrix $\delta_{k,k'}$
and an off-diagonal component, from which we extract the largest
eigenvalue $\lambda$ and Eigenvector  $U_{\lambda}(k)$.
At the next iteration step, we model the diagonal component
 as $\delta_{k,k'}(1 - \lambda U_{\lambda}^{2}(k))$,
subtract that quantity from the original matrix, and extract the principal Eigenvector of
the result.
We repeat this process until the difference between the original and the
model converges -- four times in this case.
At the end of the iterative step $i$, the parameterized matrix is thus given by
\begin{equation}
    \rho^{i}(k,k^\prime) = \delta_{k,k'} \left( 1 - 
      \lambda^{i-1} [U_{\lambda}^{i-1}(k)]^{2}\right)+ 
           \lambda^{i} U_{\lambda}^{i}(k) U_{\lambda}^{i}(k')
    \label{eqn:eigen_init}
\end{equation}
After the last iteration, we update the first term with the latest $\lambda$
and $U_{\lambda}(k)$.
The diagonal component typically starts off as unity in the largest scales and
fades away in the non-linear regime.
In this case, the strongest eigenvalue is almost two orders of magnitude larger
than all others, hence we keep only a single Eigenvector 
(see \citep{Harnois} for the application of this method when
the angular dependence of the covariance matrix is preserved).

The modeled covariance matrix is then recovered from this noise-reduced cross-correlation
matrix with the diagonal elements of the original covariance $C(k,k)$
(not to be confused with the diagonal component of the factorization of $\rho(k,k')$) following \eqn{corrmatrix}.
Although we do need $N^2$ matrix elements to start with, the power of this decomposition
relies on the fact that the prior is accurate at the few percent level, and that everything that does not
fit into this form is considered as noise.
We thus recover smooth and accurate covariance matrices, even though the input elements are noisy.
The explanation for this is that one needs only to measure the N diagonal elements of the covariance matrix
(which are generally the easiest to resolve) plus one Eigenvector
(which combines the data of a large portion of the matrix into another N elements).

\begin{figure}
        \centering
        \includegraphics[width=3.5in]{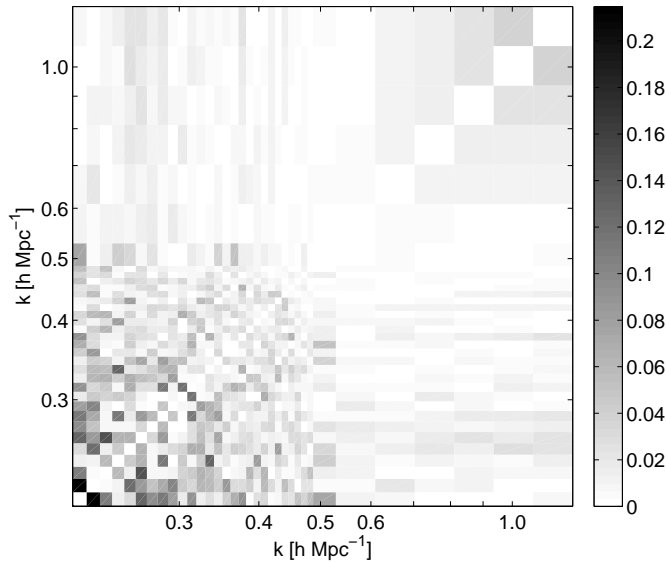}
        \caption{Absolute value of the fractional error between the modeled 
                    covariance matrix and the original,
        from unreconstructed density fields at $z=0.5$. 
        As expected, the lower $k$ modes are more noisy due to lower statistics.
        We still achieve an agreement  of about
        ten percent at $k \sim 0.22\ \rmn{Mpc}^{-1}h$, and even better for 
                smaller scales. }
        \label{fig:frac_error}
\end{figure}

We present the  fractional error between the modeled and  the original matrices
in  \fig{frac_error},  and observe that individual elements match the model
at the ten percent level at $k\sim 0.2\ \rmn{Mpc}^{-1}h$. This improves to the
few percent level for even smaller scales.
We expect these higher $k$ modes to be more accurately measured to start with,
since they come from
the angle averaging of  $k$-shells that contain  much larger number of cells.

\begin{figure}
        \centering
        \includegraphics[width=3.5in]{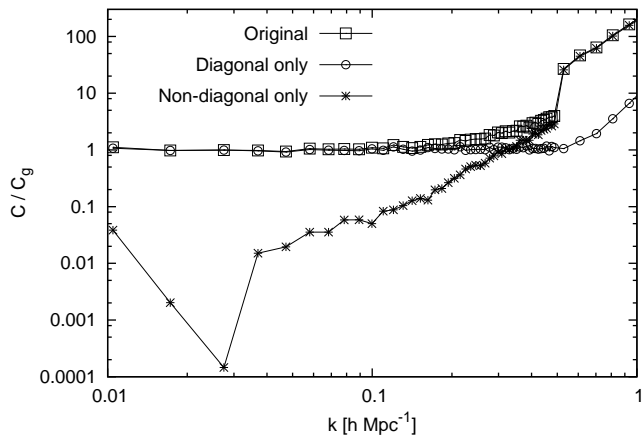}
        \caption{The ratio of the diagonals to the Gaussian approximation. 
	The squares are from the original matrix, the open circles are from 
           the diagonal component of the
        factorization, and the stars are from the diagonal of the off-diagonal 
           contribution. 
	The diagonal component is closer to the Gaussian prediction down to 
            much smaller scales,
        as most of the non-Gaussianities are factorized in the other term.}
        \label{fig:ratio_to_gauss}
\end{figure}

\begin{figure}
        \centering
        \includegraphics[width=3.5in]{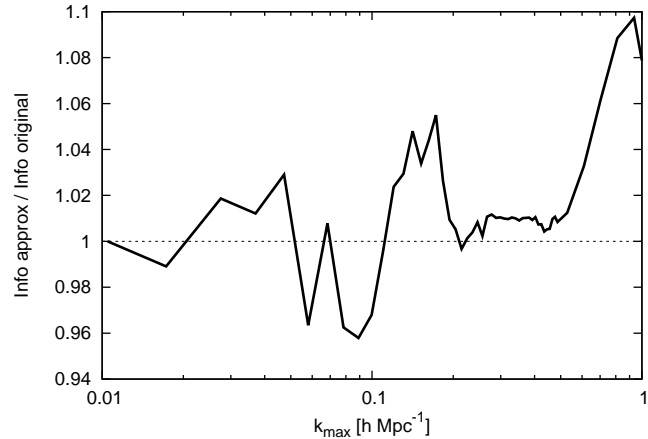}
        \caption{The ratio between the Fisher information of our modeled 
        covariance matrix 
        and that of the original, from unreconstructed density fields at 
        $z=0.5$.
        The departure from the original is at most $10\%$, and the agreement is
        otherwise
	at the few percent level.}
        \label{fig:info_ratio_model}
\end{figure}

We then present in \fig{ratio_to_gauss} the diagonals of the original 
covariance matrix
and of both components, divided by the Gaussian prediction. 
The sum of the two components is equal to the original matrix on the diagonal 
by construction.
The off-diagonal components of the factorization can be tested against the 
original
by comparing the cumulative Fisher informations (\eqn{info}).  
We observe that the information from the factorized covariance matrices
exhibit the same essential features, namely a Gaussian increase up to about 
$k \sim 0.2\ \rmn{Mpc}^{-1}h$,
followed by  a saturation plateau.
We present the ratio of this modeled information to the original distribution 
in \fig{info_ratio_model}
and observe that the factorization reproduces the information content at the 
few percent level.
Although the figures presented in this section correspond to the 
unreconstructed densities at $z=0.5$,
we achieve comparable performances on reconstructed density fields at all 
measured redshifts. 

This factorization has one extra advantage, which comes from the fact that 
$ U(k)\sqrt{C(k,k)}$ is binning independent. 
The original covariance matrix is binning dependent, since the number of modes
entering each element varies as we change  from linear to logarithmic bands, 
for example. 
This is correct, but it has a major inconvenience -- the cross-correlation 
coefficient matrix (\eqn{covariance}) 
visually changes significantly. It becomes a tedious task to compare figures 
from different authors.
We can therefore attempt to fix this problem by constructing binning 
independent quantities.

The diagonal component, $1-\lambda U^{2}(k)$, is close to Gaussian, as seen in 
\fig{ratio_to_gauss},
hence it is roughly inversely proportional to the number of modes in the bin. 
We scale it with the ratio between the measured number of modes and the 
continuous limit case: 
$N_{cont} \sim 4/3 \pi (k/k_{min})^{3}$, and obtain a binning independent 
quantity. 
We have also shown in this section that 
the off-diagonal elements of the covariance matrix are well modeled by 
$\lambda U(k)U(k')\sqrt{C(k,k)C(k',k')}$, which are binning independent as well.
The solution is thus to replace the diagonal 
of the original covariance matrix such that
\begin{equation}
C(k,k) \rightarrow \left[1-\lambda U^2(k)\right] \frac{N}{N_{cont}} + 
\lambda U^{2}(k). 
\end{equation}
This bin independent result is made available thanks to the factorization presented above,
which allows us to isolate the bin dependence (that comes exclusively from the diagonal component) and to apply our correction.
The cross-correlation coefficient matrix calculated that way is shown in 
\fig{corrmatrix},
and compares well with \cite{rh2005}, apart from the 
anti-correlation between
the largest  and the smallest modes (see section \ref{subsec:covmatrix}, 
and note that it is not clear that the simulations \cite{rh2005} were large
enough to contradict our anti-correlation result).


\section{Parameter estimation}
\label{sec:parameterestimation}

\subsection{Optimal estimator}
\label{sec:optimalestimator}
For a set of parameters $p_i$, the Fisher matrix $F_{ij}$ is given by the 
inverse of a covariance matrix
\begin{equation}
	F_{ij} = C_{ij}^{-1}
        \label{eqn:Fish}
\end{equation}
where $C_{ij}\equiv C(k_i,k_j)$ or $ C_{g}$, as described in the previous 
sections. Since our 
$C_{ij}$ is derived from power spectra, the parameters in $F_{ij}$ can be 
thought of as the power for each $k$ mode in the power spectrum. 
 To obtain the 
Fisher matrix for another set of parameters $q_\mu$, we can project the power 
spectrum Fisher matrix onto the new parameter space using
\begin{equation}
	F_{\alpha\beta} = \sum_{k_i,k_j}^{k_{max}}\partiald{P(k_i)}{q_\alpha} 
F_{ij} 
\partiald{P(k_j)}{q_\beta}.
	\label{eqn:fisher}
\end{equation}
Using $F_{\alpha\beta}$, the optimal error estimator for parameter $q_\alpha$ 
is simply 
\begin{equation}
	\Delta q_\alpha = \left( F^{-1} \right)_{\alpha\alpha}^{1/2}.
	\label{eqn:optimalestimator}
\end{equation}
The estimators for cases (ii) and (iii) are obtained from substituting 
\eqns{covariance}{covariancegauss}
respectively in \eqn{Fish}.

For our purposes, since we are interested in estimating the fractional errors of $s$ (1D) or $s_\perp$ and 
$s_{||}$ (2D), we set $q\equiv \ln s$ in 1D, and $q_1\equiv \ln s_\perp$ and 
$q_2\equiv \ln s_{||}$ in 2D. The derivatives in \eqn{fisher} can be evaluated 
in a number of ways. We can either produce a variety of $P(k)$ using different 
parameters $s$ and take their finite differences, or we can 
parametrize $P(k)$ as a function of $s$, and evaluate the 
derivatives analytically. In both cases, we decompose the power spectrum into 
a smooth component and a wiggly component
\begin{equation}
	P(k) = P_{smooth}(k) + P_{wiggle}(k).
\end{equation}
When using BAO to measure distances, all information is manifested as 
``wiggles'' in the power spectrum. To ensure that all our measurements 
originate 
from BAO, we subtract $P_{smooth}(k)$ \citep{EH1998} from 
our full $P(k)$ prior to taking derivatives. For the rest of this paper, we 
refer
$P_{wiggle}(k)$ and $dP_{wiggle}/dx$ to simply $P(k)$ and $dP/dx$ for brevity.

The finite differencing of $P(k)$ is done by 
producing many $P(k)$ using {\sc CMBFAST}\footnote{http://cmbfast.org/}, 
and dividing their differences by the differences of $s$ being used. On the 
other hand, readers familiar with \citet{se07} would recall that they suggested
an analytical form of the wiggles using a damped sync function, and they 
approximated
the square of the sinusoidal components in the Fisher matrix as a constant. 
In our work,
since we use non-Gaussian (hence non-diagonal) covariance matrices, we do
not adapt that constant approximation, as the detailed structure of the signal 
becomes important once off-diagonal errors are present.
Nevertheless, it is interesting to show what effect this approximation and
the off-diagonal elements of the covariance matrix can have on our results.
See Appendix A for this discussion.

\subsection{Sub-optimal estimator}

In a data analysis, non-linear covariance matrices are usually hard to measure 
with a high 
signal to noise, especially in surveys that exhibit complex selection functions.
What is often done in that case is to assume Gaussianity in the data, 
while ignoring the fact that the band powers themselves are actually 
non-Gaussian. 
The values extracted for the mean and the error on the BAO dilation scale are 
not properly weighted,
since they assume that all errors in $k-$bands are uncorrelated. As mentioned 
earlier,
the resulting mean is sub-optimal, while the error bars are most likely off by 
an unknown amount.
Since we now have measured a non-linear covariance matrix $C$, we are in a position
to compare 
the correct error bars for existing data analysis (case (i))
with those quoted in literature, which approach case (iii) at various levels. 
We recall that the difference between sub-optimal and optimal is somewhat 
reduced for analyses that 
model the non-linearities in the fields. 

We derive a ``sub-optimal'' estimator by solving the linear system $Ax=b$
where $x$ is a vector  containing a set of cosmological parameters of interest
(here we consider only  $x\equiv \Delta \ln s $, hence $x$ is a scalar,  
but this method could be generalized to include  $\ln s_{\perp}$ and $\ln s_{||}$ for instance), 
and $b$ is a noisy observable, associated with a noisy covariance matrix $\tilde{C}$. 
In our case, the observable under study is $\Delta P(k)$, the deviation from the mean 
of the power spectrum.
With this correspondence, we get $A\equiv\partial{P(k)}/\partial \ln s$ -- a vector in our case.
To estimate each component of $x$, we first weight each observed point by the inverse
of the  covariance matrix associated with the observation of $b$, i.e. $\tilde{C}$, 
and then  proceed to solve for $x$ by taking pseudo-inverses such that
\begin{equation}
	x = (A^T \tilde{C}^{-1} A)^{-1}A^T \tilde{C}^{-1}b.
\end{equation}
Finally, the errors on the elements of $x$ are given by the diagonal components of the 
covariance matrix $\langle x^2\rangle=x x^T$. 
We obtain the following estimator for the error in $\ln s$:
\begin{equation}
	\langle (\Delta \ln s)^2\rangle = (P_s^T \tilde{C}^{-1} P_s)^{-1} P_s^T \tilde{C}^{-1} C \tilde{C}^{-1} P_s(P_s^T \tilde{C}^{-1} P_s)^{-1}
	\label{eqn:gaussianestimator}
\end{equation}
where $P_s \equiv \partial{P}/\partial \ln s$, and $P_s^T C P_s$ is a
vector-matrix-vector product similar to \eqn{fisher}, and where $C=bb^T$ is the
improved estimate of the 
non-linear covariance matrix, which we obtained from our simulations.
Notice that if the true 
covariance matrix was indeed Gaussian (i.e. $C = \tilde{C} = C_g$), we would 
recover the 
optimal estimator where $\langle(\Delta \ln s)^2\rangle = (P_s^T C_g^{-1} 
P_s)^{-1}=F_{11}^{-1}$. In other words, cases (i), (ii), and (iii) would be 
identical.
Conversely, if the original matrix was already the optimal measurement (i.e. $C = \tilde{C}$), we 
would get
$\langle(\Delta \ln s)^2\rangle = (P_s^T C^{-1}P_s)^{-1}$, i.e. cases (i)  and 
case (ii)  would be the same, possibly different from case (iii).
We also note that the {\it inverse} of the true covariance matrix is not involved 
in \eqn{gaussianestimator}.

In the next sections, we consider the case where Gaussianity was originally 
assumed in a BAO analysis, 
we correctly estimate the error of $x$ (with this sub-optimal estimator)
and compare the results produced with an optimal estimator.

\subsection{Effects of reconstruction}
\label{sec:effectsofreconstruction}

As seen in \fig{xi}, density reconstruction  sharpens the acoustic peak in 
the correlation function. This is equivalent to reducing the non-linear damping 
of the wiggles in the power spectrum. We parametrize the reduction of damping 
due to reconstruction by an extra factor in front of $\Sigma_{nl}$ in the 
damping factor \citep{se07} such that

\begin{equation}
	\exp\left[ -\frac{k^2 \Sigma_{nl}^2}{2} \right] \rightarrow 
                 \exp\left[ - \frac{ (1-f)^2 k^2 \Sigma_{nl}^2}{2} \right]
    \label{eqn:reconfac}
\end{equation}
where $f=1$ represents 100\% reconstruction, canceling any non-linear effects. 
In reality, such a case is unachievable, as some information has been 
irreversibly lost. 
In principle, we could measure $f$ by extracting the power spectrum wiggles 
before and after 
reconstruction, and find the best fit damping factor. Looking at \fig{xi},
we see that the reconstructed correlation at $z=0.5$ is very similar to the 
correlation at $z=2$. The ratio of growth factors between these two redshifts
is 0.55, and the reconstructed field actually looks slightly better than the 
$z=2$ field, so we adopt the 
standard $f=0.5$, which reduces non-linear damping by 50\%
\citep{se07, chime}.


\section{Results and discussion}
\label{sec:resultsanddiscussion}

We set up hypothetical surveys with volume $V=(1\ h^{-1}\rmn{ Gpc})^3$ centered 
on redshifts $z=0.5$ and $z=1.0$. We then produce fractional distance errors $
\Delta s/s$ using the three aforementioned estimators:
\begin{itemize}
	\item Sub-optimal estimator (\eqn{gaussianestimator})
	\item Optimal estimator (\eqn{optimalestimator}) with a non-Gaussian 
                    covariance matrix (\eqn{covariance}),
	\item Optimal estimator (\eqn{optimalestimator}) with a Gaussian 
                    covariance matrix (\eqn{covariancegauss}).
\end{itemize}
The distance error measurements use only the information up to a limiting 
$k_{max}$
in the covariance matrix; all $k>k_{max}$ are marginalized over 
(i.e. cut off from the covariance matrix before it is inverted).
For each error estimate, we also produce bootstrap error bars to show the
convergence of our set of 1000 simulations. This is done by picking 1000 random 
simulations (allowing repeats) from our set, and taking the standard deviations
of the results using 2000 such random sets. 
The noise reduction technique is performed for every set.


The only redshift dependence in the estimator  comes from the non-linear 
damping scale 
$\Sigma_{nl}(z)$ (\eqn{reconfac}).
 In addition, we know from \figs{corrmatrix}{information} that the covariance 
matrices in the linear regime 
are similar to the Gaussian covariance matrix. Therefore, we expect errors at 
small $k$ 
to be hardly distinguishable among estimators and redshifts.
At large $k$, however, we expect the effects of the covariance matrices and the derivatives to have a  
more pronounced effect on the estimator. The redshift dependent damping scale 
$\Sigma_{nl}(z)$ 
becomes important and distinguishes between $z=0.5$ and $z=1.0$. 

\begin{figure}
    \centering
    \includegraphics[width=3.5in]{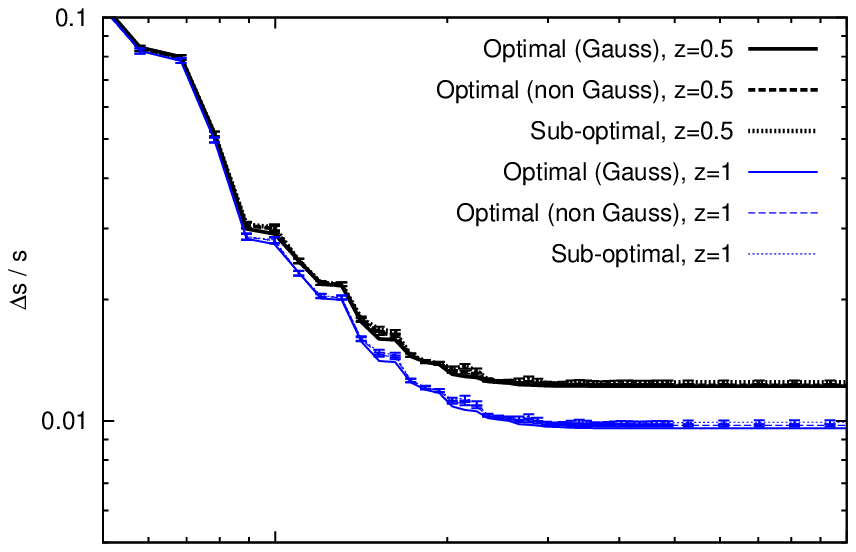}
    \includegraphics[width=3.5in]{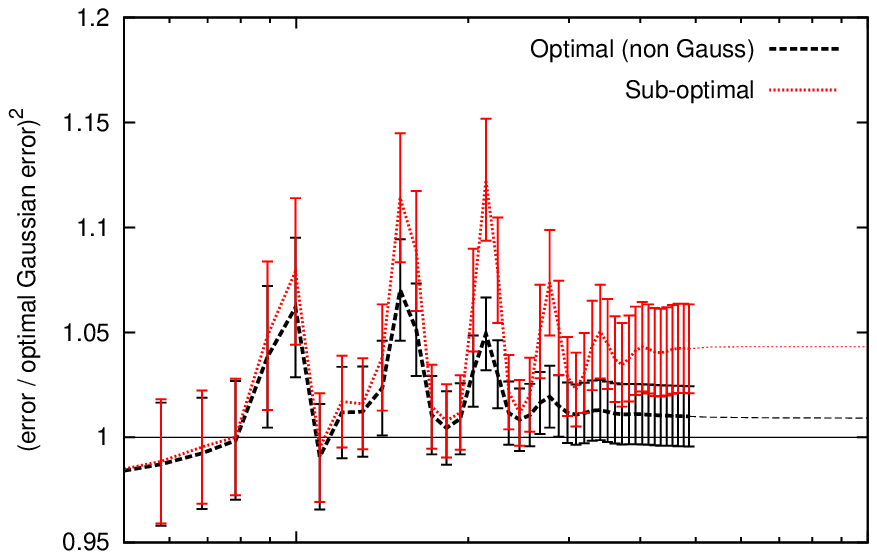}
    \includegraphics[width=3.5in]{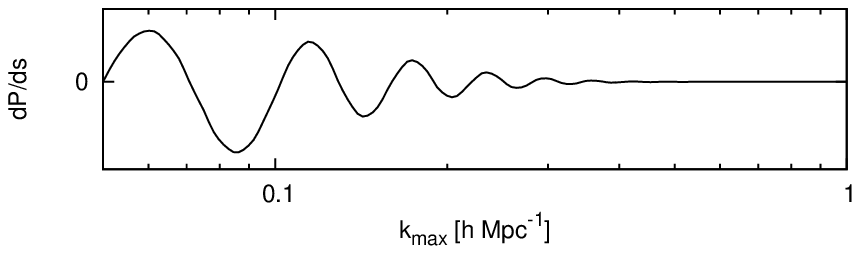}
    \caption{{\it Top panel}: Fractional distance errors using optimal and 
           sub-optimal estimators, without reconstruction.
{\it Optimal (Gauss)} and {\it Optimal (non Gauss)} denote Fisher estimators 
   from covariance matrices $C_g$ and $C$ respectively. 
{\it Sub-optimal} denotes the correct non-Gaussian errors if Gaussianity was 
assumed.
The error bars on each line are the 1-$\sigma$ bootstrap error bars for the 
statistical fluctuations
in our set of simulations. {\it Middle panel}: Optimal 
non-Gaussian and sub-optimal errors, each divided by the optimal Gaussian 
errors at $z=0.5$. This panel shows
that each error estimate deviates by only a few percent from the Gaussian 
estimate. {\it Bottom panel}: Our $dP/ds$ template which shows that
the regime $k>0.5$ Mpc$^{-1}h$ is no longer relevant for BAO analysis, as the 
wiggles are mostly damped
and converged to zero.
}
    \label{fig:ds_kmax}
\end{figure}

The top panels of \figs{ds_kmax}{ds_kmax_recon} show the measured distance 
errors vs limiting $k_{max}$, 
with and without reconstruction, respectively.
The bottom panels shows the squares of the ratio of cases (i) and (ii) to case 
(iii) from the top plot at $z=0.5$.
In all scenarios, we consider only the modes $k\lesssim 0.5$ Mpc$^{-1}h$ where 
our simulations
remain reliable (\fig{powerspectrum}). Moreover, we do not expect modes of $k>0.5$ Mpc$^{-1}h$ to 
carry any BAO information, as the wiggles on small scales suffer from Silk 
damping \citep{EH1998} as well as
non-linear damping (\sect{effectsofreconstruction}).

\begin{figure}
    \centering
    \includegraphics[width=3.5in]{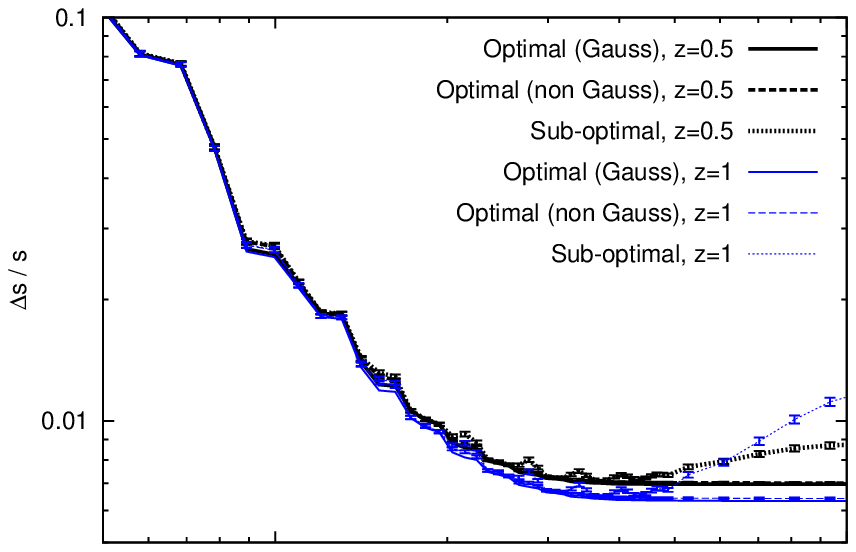}
    \includegraphics[width=3.5in]{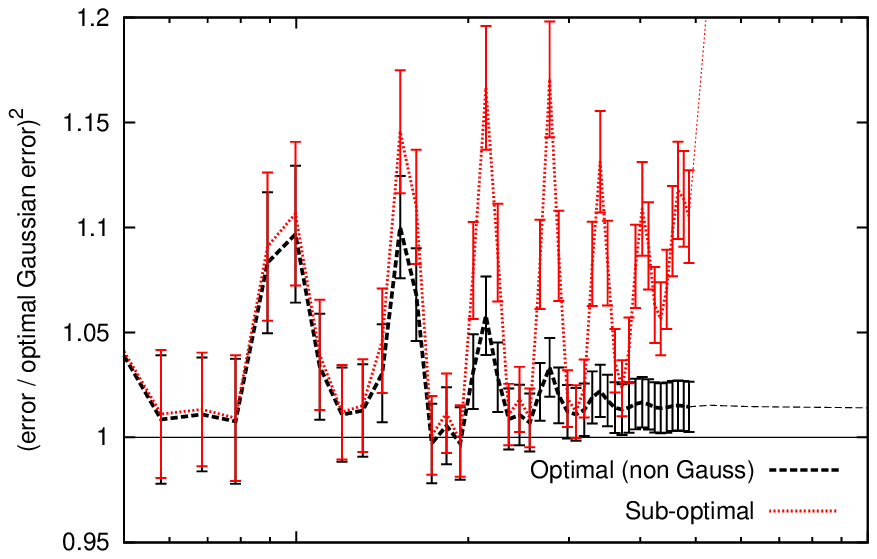}
    \includegraphics[width=3.5in]{bootstrapbefore/dPda.eps}
    \caption{Fractional distance errors after reconstruction. Compared to 
\fig{ds_kmax}, this now includes reconstructed covariance 
matrices and a reconstruction factor of $f=0.5$. In the sub-optimal case, the
increase at the regime $k>0.5$ Mpc$^{-1}h$ is a statistical curiosity, but 
irrelevant for a BAO analysis. We base our conclusion on the values for 
$k\lesssim0.5$ Mpc$^{-1}h$. This cut-off coincides with our choice to switch
from linear binning to logarithmic binning, but we have checked that the 
above plots contain no binning artifacts.
}
    \label{fig:ds_kmax_recon}
\end{figure}

\fig{ds_kmax_recon} shows the distance errors with $f=0.5$ and covariance 
matrices from reconstructed simulations. Since some of the wiggles are indeed
recovered, the optimal distance errors decreased by 50\% to 70\% compared to the case with no reconstruction (\fig{ds_kmax}).
More importantly, though, the discrepancy between sub-optimal and optimal 
estimates becomes slightly more severe. In the presence of reconstruction, a 
naive Gaussian assumption can underestimate the variance of the errors by up 
to 20\% near
the trans-linear scales ($k \sim 0.2$ Mpc$^{-1}h$). Somewhat disturbingly, at 
small scales $k>0.5$ Mpc$^{-1}h$ the sub-optimal
estimate deviate significantly from the optimal estimate. A sub-optimal 
estimator certainly is not expected to provide
the same result as an optimal estimator. And also, as mentioned above, the 
regime $k>0.5$ Mpc$^{-1}h$ is irrelevant to
BAO analysis. Therefore, we did not investigate this behavior further.


Our results from \fig{ds_kmax} show that the three treatments of the errors
have similar constraining power on the BAO dilation scale.
This conclusion  is consistent with that from \citet{jap},
which measured that cases (ii) and (iii) give almost identical results, without
reconstruction. 
However, we push the envelope further and show that, first, the sub-optimal
estimator also behaves similarly, with deviations by up to $15\%$.
Second, we show that the reconstruction of density fields
improves the constraints on dilation by about 70\% at $z=0.5$. 
The improvement is more modest at higher redshifts, where additional
BAO peaks are still in the linear regime.
Third, we conclude that with reconstruction, Gaussian assumption can 
underestimate
the errors on the dilation scale in a similar level to the case without 
reconstruction.

\section{Conclusion}
\label{sec:conclusion}

We have addressed some aspects of the problem of measuring non-Gaussian error 
bars in a 
BAO analysis. We have investigated the full optimal quadratic non-Gaussian 
estimator 
on reconstructed density field, and quantified the significance of the 
non-Gaussianities on the BAO dilation scale error.
A major subtlety is that the optimal estimator requires an accurate measurement 
of the inverse of the covariance matrix of power spectrum, and the actual 
uncertainty
on that inverse had never been measured. The accuracy of the inverse is 
generally bin
dependent, and convergence requires a large number of N-body simulations.
We have overcome this problem with a factorization technique that involves
an iterative eigenvalue decomposition of the covariance matrix,
which we measured from 1000 N-body simulations.
We further measured the uncertainty of the inverse of the matrix with bootstrap
re-sampling, and were able to achieve convergence at the percent level.

Having confidence in the inverse matrix, we quantitatively compared the 
measurement of the error on the BAO dilation factor obtained
with different estimators.
We construct an {\it optimal estimator}, which
gives the most accurate measurement of the error achievable on the BAO dilation
scale. 
It is derived from a covariance weighted Fisher matrix, which is 
constructed out of the inverse of the non-linear power spectrum covariance 
matrix.
We compared our results with those obtained from the purely Gaussian forecast,
and we measure significant discrepancies of up to ten percent in the error on 
the dilation scale. 

We also measured non-Gaussian error bars on the mean BAO dilation scale that 
has been 
obtained with a Gaussian estimator, as is usually encountered in the literature.
Because we have confidence in the accuracy of our covariance matrix, 
this {\it sub-optimal estimator} provides a robust estimate of the
error bars on the BAO dilation scale.
To illustrate our point, we considered the case where the original dilation
scale was measured under standard Gaussian statistics.  We found that the
variances of those measurements can differ by up to 15\%, compared to our 
optimal estimator.
Many data analyses did include non-Gaussianities in their BAO error estimator, 
hence the discrepancy between these and our optimal estimator 
is likely to be more modest than that obtained in this work. 

We note in passing that these results were entirely obtained from N-body 
simulations, 
hence the effect of the survey selection function has been factored out of our 
problem.
Constructing optimal estimators with actual data will however be more 
challenging,
since it has to include such an effect, which effectively couple Fourier modes
from different bins, in addition to account for the effect of bias between the 
sampled
tracers (i.e. galaxies or  21cm structure) and the underlying matter density.

We have also implemented a density reconstruction algorithm, which
recovers some of the lost BAO information due to non-linear
gravitational collapse at late times. In that case, the error on the
dilation scale is reduced by a factor of about 70\% at low redshift,
but the discrepancy between the sub-optimal and optimal estimates remains
similar to the case without reconstruction (20\% and 15\%, with and
without reconstruction). 

We mention in conclusion that in a survey, the increase in variance 
we observed when using a sub-optimal estimator is equivalent
to losing about the same percentage of survey volume, because the
variance of measurements is inversely proportional to volume.  These
discrepancies should be taken seriously into account especially when
forecasting performances of future telescopes, where the objective is
to reach percent level precision on cosmological parameters.

\section*{Acknowledgments}

The authors thank Kiyoshi Masui for valuable help and advice,
and acknowledge NSERC for their financial support.
The simulations in this work were produced on the Sunnyvale cluster at CITA.

\appendix
\section[]{Power spectrum derivative}
\label{sec:powerspectrumderivative}

Interestingly even though the Fisher information (\fig{information}) of Gaussian
and non-Gaussian covariance matrices differ by about an order of magnitude near
$k\gtrsim 0.2$ Mpc$^{-1}h$, the measurement errors (\figs{ds_kmax}{ds_kmax_recon})
differ by only a few percent. The reason for this is that when the oscillatory
power spectrum derivative $dP/d\ln s$ is multiplied into the covariance matrix
to compute the Fisher matrix (\eqn{fisher}), the off-diagonal elements of the
covariance matrix can be canceled out.

In \citet{se07}, they considered the regime $k\gtrsim 0.05$ Mpc$^{-1}h$ and
approximated a $\cos^2(ks)$ term, which originates from $dP/d\ln s$,
as simply $1/2$ without any oscillations when computing
their Fisher matrix. Since we are considering a similar regime, we also 
attempted
this approximation. We emphasize that this approximation is not applicable to 
us,
as our non-Gaussian covariance matrix is not diagonal. Nevertheless, it 
illustrates
the effect that an oscillatory $dP/d\ln s$ can have on distance measurements.

\begin{figure}
    \centering
    \includegraphics[width=3.5in]{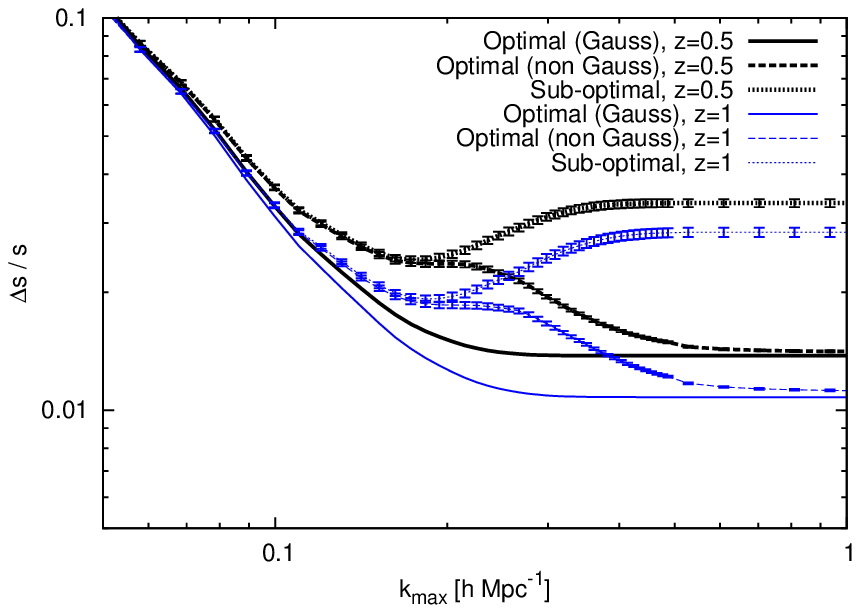}
    \label{fig:ds_approx}
    \caption{Fractional distance errors using various estimators, similar to 
                the top panel of \fig{ds_kmax}
    (without reconstruction). The only difference here is that the power 
                           spectrum derivative
    in the Fisher matrix is an approximation which does not oscillate. Compared
                  to \fig{ds_kmax},
    the {\it Optimal (Gauss)} case is similar, but the other cases are clearly 
                            different.}
\end{figure}

\fig{ds_approx} shows $\Delta s/s$ as a function of limiting $k_{max}$, using 
the
approximated derivative. This can be compared to the top panel in 
\fig{ds_kmax},
where the only difference is that \fig{ds_approx} uses the approximated 
$dP/d\ln s$,
and \fig{ds_kmax} computes $dP/d\ln s$ by finite differences of actual power 
spectra.
In both figures, the optimal estimators follow the inverse of the Fisher
information (\eqn{info}). For the optimal Gaussian case, the errors using either
forms of derivatives give similar results (other than some oscillations at 
small $k$).
This is expected since the Gaussian covariance matrix is diagonal 
(\eqn{covariancegauss}),
so the approximation is valid. For the sub-optimal cases we considered in this 
paper, however, the discrepancies
from the optimal Gaussian case reach factors of 2 to 3, depending on redshift.
This can be attributed to the fact that all elements of the covariance matrix are given a 
uniform weight. When using the finite difference derivatives, however,
different elements are weighted according to the values of $dP/d\ln s$ at their corresponding $k$ modes.   
This effectively cancels most of the contributions from the off-diagonal 
elements of the
covariance matrix, which explains why the optimal and sub-optimal errors differ by
only a few percent in that case.

\bsp

\label{lastpage}

\end{document}